\documentclass[preprint,showpacs,preprintnumbers,amsmath,aps,pra]{revtex4-1}

\usepackage{amssymb,amsfonts,amstext,graphics,graphicx,subfigure}
 
\usepackage{bigints}

\font\fb=rsfs10

\def\Qf{\begin{text}{\fb Q}\end{text}}  

\def\ntan{||\bfX_\si||}

\def\ys{Y_{\sigma}}
\def\xs{X_{\sigma}}
\def\rt{X^{2} + Y^{2}}
\def\rts{X_{\sigma}^{2} + Y_{\sigma}^{2}}
\def\dxf{\ys \frac{\delta F}{\delta X} - \xs \frac{\delta F}{\delta Y}}
\def\fq{\frac{\dxf}{\rts}}

\def\oi{\oint d\sigma}

\def\tq{\tilde q}

\def\cala{\mathcal{A}}

\def\bq{\begin{equation}}
\def\eq{\end{equation}}
\def\bqy{\begin{eqnarray}}
\def\eqy{\end{eqnarray}}

\def\vX{{\bf X}}
\def\o{\over}

\def\al{\alpha}

\def\de{\delta}
\def\De{\Delta}

\def\et{\eta}

\def\Ga{\Gamma}

\def\la{\lambda}

\def\om{\omega}
\def\Om{\Omega}
\def\ph{\phi}

\def\si{\sigma}

\def\bfx{\mathbf{x}}
\def\bfX{\mathbf{X}}

\def\vxp{{\bf x'}}

\def\p{\partial}

\def\del{\nabla}

\def\half{\frac12}
\def\intint{\int\! \int}

\def\calC{{\cal C}}

\def\ddq#1{{\delta #1\over\delta q}}
\def\pb#1{\{#1\}}

\def\ppt{{\partial\over\partial t}}
\def\ppx{{\partial\over\partial x}}
\def\ppy{{\partial\over\partial y}}

\def\del{\nabla}

\def\p{\partial}

\def\x{{\bf x}}

\def\u{{\bf u}}

\def\k{{\bf k}}

\def\bfxi{\xi\kern -4.5pt\xi}

\def\intint{{\int\negthinspace\negthinspace\negthinspace\int}}
\def\half{{1\over 2}}
\def\ltsim{\raise 0.6ex \hbox{$<$}\kern -0.7em\lower 0.6ex\hbox{$\sim$}}

\def\hr{\hbox to \hsize{\hrulefill}}

\def\Del{\nabla\kern-7pt\nabla}

\input epsf
\edef\tildex{\string~}
\let\@\tildex%
\font\sc=cmcsc10

\newcount\eqnnum
\newcount\fignum
\def\fig{\the\fignum\global\advance\fignum by 1}
\def\figs{\the\fignum}

\def\section#1{\medskip\goodbreak\noindent{\bf #1}\medskip}
\def\subsection#1{\medskip\goodbreak\noindent{\sl #1}\medskip}
\def\subsubsection#1{\medskip\goodbreak\noindent{\sc #1}\medskip}

\def\references{\advance\leftskip\parindent\parindent=-20pt}

\def\calC{{\cal C}}

\def\genfigr#1#2#3#4{
\vbox to #1 true in{
\includegraphics{#3} 
}
{\medskip{\narrower{\noindent #4\smallskip}}\medskip}
}
\def\genfigrp#1#2#3#4{
\vbox to #1 true pt{
\includegraphics{#3} 
}
{\medskip{\narrower{\noindent #4\smallskip}}\medskip}
}
\newcount\figsize
\newcount\figoffset
\newcount\figpt
\def\newfig#1#2#3{
\figsize=12
\figoffset=5
\multiply\figsize by #1
\multiply\figoffset by #1
\genfigr{#1}{\the\figsize}{#2 voffset=\the\figoffset}{#3}
}
\def\newfigp#1#2#3{
\figsize=#1
\figoffset=#1
\divide\figsize by 6
\divide\figoffset by 14
\genfigrp{#1}{\the\figsize}{#2 voffset=\the\figoffset}{#3}
}

\def\fig#1#2{\medskip\epsfysize=#1 in\centerline{\epsfbox{#2}}}

\begin{document}

\title{Jovian vortices and jets}
 
\author{Glenn R. Flierl}
\affiliation{Department of Earth, Atmospheric and Planetary Sciences,\\ 
MIT, Cambridge, MA}
\email{glenn@lake.mit.edu}
\author{ Philip J. Morrison}
\affiliation{Department of Physics and Institute for Fusion Studies,\\  
University of Texas, Austin, TX}
\email{morrison@physics.utexas.edu}
\author{Rohith Vilasur Swaminathan}
\affiliation{Department of Earth, Atmospheric and Planetary Sciences, MIT, Cambridge, MA (currently in  India)}
\email{rohith.vs@gmail.com}

\begin{abstract}
We explore the conditions required for
isolated vortices to exist in sheared zonal flows and the stability of the
underlying zonal winds. This is done using the standard 2-layer
quasigeostrophic model with the lower layer depth becoming infinite;
however, this model differs from the usual layer model because the lower
layer is not assumed to be motionless but has a steady configuration of
alternating zonal flows \cite{stamp93}. Steady state vortices are obtained
by a simulated annealing computational method introduced in \cite{vallis89},
generalized and applied in \cite{pjmF11} in fluid flow, and used in the
context of magnetohydrodynamics in \cite{pjmF17,pjmFWI18,pjmBKM18}. Various cases of
vortices with a constant potential vorticity anomaly atop zonal winds and
the stability of the underlying winds are considered using a mix of
computational and analytical techniques.
\end{abstract}

 \maketitle

\section{{\bf1.}  Introduction}

A great red spot on Jupiter has been observed for centuries, with its present
manifestation dating back to telescopic observation in 1830.  With the
technological observational advancements and spacecraft measurements of modern
times, additional Jovian vortical features have been discovered as well as such
features in other planets. Many theoretical ideas have been proposed, yet it 
seems at minimum the red spot is a vortex enmeshed in zonal flow. 

In  this paper we investigate the conditions required for isolated vortices to
exist in sheared zonal flows and the stability conditions for the maintenance
of zonal winds.  To this end we use the simple model of \cite{dowling89} to
illustrate the basic concepts and to explore the sizes and shapes of the
vortices.  The model  can be viewed as a version of the standard 2-layer
quasigeostrophic (QG) model with the lower layer depth becoming infinite.
Unlike the more common $1\frac{1}{2}$-layer model, the lower layer is not assumed
to be motionless but rather have a steady zonal flow, $U(y)$.  Since vortex
stretching produced in the lower layer by vertical movement of the interface
is negligible, this flow can remain unchanged.  We will call this a
$1\frac34$-layer model to distinguish it from either the motionless deep layer 
case or the one in which the deep layer evolves.  

In Section 2 we will review how the $1\frac{3}{4}$-layer model emerges from the 2-layer model. Also in this section  we review the noncanonical Hamiltonian formalism \cite{pjm82,pjm98,pjm05} and briefly describe the Dirac bracket formalism,  a Hamiltonian technique for the imposition of constraints.  Both formalisms will be used later when we describe  the  simulated annealing (SA) procedure  for obtaining steady states  \cite{vallis89} and our generalization, the Dirac bracket simulated annealing (DBSA) procedure,  introduced in \cite{pjmF11}.

Since the jets we consider can have $\beta-U_{yy}$ changing sign, by the Rayleigh criterion 
they may be  subject to rapid instability and break-down in either the $1\frac{1}{2}$-layer
model or standard 2-layer model.  We will re-examine the idea that steady deep flow
can stabilize the jets \cite{stamp93} using the energy-Casimir method and
relate it to the existence of isolated vortices in Section 3. Furthermore, in
\cite{kaspi07} it was shown that the ``two-beta'' model reduces the growth
rates significantly and limits the unstable waves to small scales.  In this
system, the deep fluid has a strong reverse $\beta$ effect because of large
vertical extent of the flows in the direction parallel to the rotation vector,
which is appropriate when the entropy gradients are small \cite{yano94}.

In Section 4 we describe steady states composed of localized vorticity
anomalies, vortex patches, embedded in the layer zonal flows.  From the
equations for the dynamics relative to the jets we obtain integral conditions
that are useful for identifying the allowed positions of the anomalies.
Importantly, we obtain an artificial dynamics that is adaptable to a contour
dynamics version of the simulated annealing technique \cite{pjmF11}, which we
will subsequently use to explore vortices in sinusoidal shear flows.

Section 5 begins with a description of the Hamiltonian structure for contour
dynamics in the usual incompressible two-dimensional Euler context.  This
structure, which applies to nonsingle-valued contours, was described briefly
in \cite{pjm05}, but is due to two of us (GRF and PJM) and has been used and
described in talks over the past 25+ years.  Next we describe the DBSA  technique in the context of  contour dynamics,  then, as a warmup, show how to use it to construct the Kirchhoff ellipse and its generalization with a background shear.  Lastly in Section 5, we describe constraints needed for application of DBSA to Jovian vortices.

In Section 6 a variety of Jovian vortices on jets are constructed by the contour dynamics DBSA technique with Dirac constraints chosen to be the linear momenta. The case with $\beta=0$ and a  depth independent background flow is  first considered.  The parametric dependence of vortices centered at $y=0$, where the background flow reverses,  is investigated.  Using DBSA with  $\beta\neq 0$, a vortex initially centered at $y\neq 0$ relaxes to a state consistent with the constraints.  A triangular shape reminiscent of  observations of Jupiter is seen to naturally emerge.  Lastly in this section,  a comparison of the contour dynamics solutions to DBSA applied to the full $1{3\o4}$-layer model is made.

Finally, we conclude and summarize the paper in Section 7.

\section{{\bf2.}  Model and technique}

\subsection{{\it2.A.}  Layer model review}

The standard 2-layer model of GFD \cite{pedlosky} is composed of  two coupled advection equations,
\bq
\ppt q_i + [\psi_i,q_i] = 0\,,
\qquad i=1,2\,, 
\eq 
where the potential vorticity in each layer is given by
\bq q_i = \nabla^2\psi_i + F_i(\psi_{3-i} - \psi_i) +\beta_i y\,.  
\eq 
Here  $\nabla^2\psi=\psi_{xx} +\psi_{yy}$ with subscript denoting partial derivative, 
$[\psi,q\,]=\psi_xq_y-\psi_yq_x$ is the Jacobian or ordinary bracket operator,  and $i=1$
corresponds to the upper layer. This formulation allows for two stretching
terms, $F_i$, and (non-standard) two values for $\beta_i$.  For convenience we
let $F_1=F$ and $F_2= \de F$, whence $\de=H_1/H_2$ is the ratio of the
respective heights of the two layers.

The standard $1{1\o2}$-layer model is obtained by letting $H_2\rightarrow
\infty$ keeping $H_1$ fixed, with the additional assumption that $\psi_2\equiv
0$.  However, in this limit, the bottom layer may remain dynamic with a
vorticity $q_2=\nabla^2 \psi_2+ \beta_2y$ that evolves independent of the
upper layer.  If the lower layer resides in a steady state, say $\psi_2(\x)$,
then we arrive at the $1{3\o4}$-layer model, with the  upper layer dynamics
in the presence of a steady deep layer flow governed by 
\bq 
\ppt q + [\psi,q]
= 0
\label{upper}
\eq
with 
\bqy
q &=& (\del^2-F)\psi+F\psi_2(\x)+\beta y
\nonumber\\
&=& (\del^2-F)\psi+T(\x)\,,
\label{pv175}
\eqy where without risk of confusion we have dropped the upper layer subscript
$1$.  In Section 4 we will modify the dynamics of this $1{3\o4}$-layer model
of \eqref{upper} and \eqref{pv175} in such a way as to make it amenable to
contour dynamics, which we will use subsequently in our analyses.
Beforehand, let us now turn to the Hamiltonian description possessed by the
model of \eqref{upper} and \eqref{pv175}.

\subsection{{\it2.B.} Hamiltonian structure of $1{3\o4}$-layer model}

The model is a Hamiltonian field theory with its Hamiltonian functional $H$
naturally being the total energy, kinetic plus potential, written as a
functional of the dynamical variable $q$: \bq H[q]=-\half\intint d\x\,d\x'\;
\big[\,q(\x)-T(\x)\big]\,G(\x-\x')\,\big[\,q(\x')-T(\x')\big]\,,
\label{ham}
\eq
with $d\bfx=dxdy$ and the Green's function, $G(\x-\x')$, satisfying
\bq
(\del^2-F)G(\x-\x') = \delta(\x-\x')
\,.
\label{green}
\eq 
Because the dynamical variable $q$ does not constitute a set of
canonically conjugate field variables, the system takes noncanonical
Hamiltonian form (see e.g.\ \cite{pjm82,pjm98,pjm05} for review) in terms of
the following Poisson bracket: 
\bq 
\pb{A,B} = \int d\x\; q(\x)
\left[\ddq{A},\ddq{B}\right]=: \int d\x\ \ddq{A}\, \mathbb{J}_{\mathrm{QG}}\,
\ddq{B} \,,
\label{PB}
\eq 
where $\mathbb{J}_{\mathrm{QG}}:= -[q, \, \cdot\, ]$, the usual Jacobian
expression defined above, is the Poisson operator.  The bracket of \eqref{PB} is a binary,
antisymmetric operator, on functionals $A,B$, expressed in terms of their
functional derivatives, $\de{A}/\de q$ and $\de{B}/\de q$.  Most importantly,
the bracket of \eqref{PB} also satisfies the Jacobi identity, 
$$ 
\pb{\pb{A,B},C}+\pb{\pb{B,C},A}+\pb{\pb{C,A},B} =0\,,
$$ 
for all functionals $A,B,C$.   The Jacobi identity is the essence of being Hamiltonian -- it guarantees the existence of a coordinate change to the usual canonically conjugate variables. With the Hamiltonian of \eqref{ham} and the bracket \eqref{PB},  the
time evolution of a functional $A[q]$ is given by 
\bq 
\ppt A = \pb{A,H}\,.
\label{PBdyn}
\eq
For example, if $A=q(\x)=\int d\x' q(\x')\delta(\x-\x')$, then $\de A/\de q =
\delta(\x-\x')$ and, from definition of $H$,
$$
\ddq H = -\int d\x' \, G(\x-\x')\big[q(\x')-T(\x')\big] = -\psi(\x)\,,
$$
Using these in \eqref{PBdyn} and \eqref{PB} yields the $1{3\o4}$-layer model of
\eqref{upper} and \eqref{pv175}.

The antisymmetry of the bracket ensures conservation of energy $d H[q]/d
t=0$, while invariants $P_\mu$ associated with other Noether symmetries satisfy
$\pb{P_\mu,H}=0$.  In particular, if $T$ is a function only of $y$, the zonal
linear momentum, 
\bq 
P=\int d\x\,yq\,, 
\label{ymom}
\eq
will be conserved:
\bqy
\frac{d P}{dt} &=&\{P,H\}= \int d\x\, q\left[y,\ddq{H}\right] 
\nonumber\\
&=& -\int d\x\, q\ppx \ddq{H} = \int d\x\,vq=0\,, 
\eqy 
where the last equality follows given either periodic conditions or
channel walls in the north and south.  In  the following we will mostly work
in a reference frame translating at speed $c$ (yet to be determined); this is
equivalent to generating the motion with the Hamiltonian $H_c=H-cP$ rather than $H$ since $\de
H_c/\de q = -\psi-cy$.  Similarly, $T$ may possess other symmetries giving
rise to the possible class of invariants 
\bq 
P_\mu[q]=\int d\x\, \phi_\mu(\x) q
\label{Pi}
\eq 
where $\phi_\mu\in\{x,y,x^2 + y^2\}$, corresponding to the momenta arising
from two possible translational symmetries and $L$, the angular momentum
arising from rotational symmetry, respectively.

Noncanonical Poisson brackets like \eqref{PB} are degenerate and have
constants of motion associated with the null space of the Poisson operator,
$\mathbb{J}$, the so-called Casimir invariants that satisfy
$$
\pb{\calC[q],B[q]}=0
$$ for all $B[q]$.  Thus, they are constants of motion for any Hamiltonian.
The bracket of \eqref{PB} and consequently the QG equations have Casimir
invariants of the form
$$
\calC[q]=\int d\x\,C(q(\x))
$$ 
with $C(q)$ an arbitrary ordinary function; commonly used examples of such
conserved functionals are the mean of the PV itself and the potential
enstrophy 
\bq 
Z[q]=\half \int\! d\x\, q^2\,.
\label{Z}
\eq

\subsection{{\it2.C.} Dirac constraints and steady states}

In our previous work \cite{pjmF11} we used simulated annealing, a technique
based on Hamiltonian structure, to obtain a variety of steady states for QG
flows.  Our work generalized a technique previously introduced in
\cite{vallis89} by adding a smoothing metric and, importantly, Dirac constraints that allow for the relaxation to a
larger class of steady states.  Starting with an initial state, the technique
produces a dynamical system that rearranges parcels of fluid conserving each
bit's potential vorticity to achieve a maximum or minimum value of the
Hamiltonian functional.  This is done via a  time-stepping of the PV by advecting
with an artificial non-divergent velocity obtained from a so-called Dirac
bracket, a generalized Poisson bracket that builds in arbitrary invariants, e.g., for two such invariants 
$C_{1,2}$ it has the form 
\bq 
\pb{A,B}_D = \pb{A,B} +{\pb{A,C_1}\pb{C_2,B} \o
\pb{C_1,C_2}} + {\pb{A,C_2}\pb{C_1,B} \o \pb{C_2,C_1}} \,,
\label{DPB}
\eq 
where $\{C_j,B\}_D=0$ for any functional $B$.  In Dirac's original work
\cite{dirac}, the bracket $\{,\}$ of \eqref{DPB} was the usual canonical
Poisson bracket; however, his construction gives a bracket that satisfies the
Jacobi identity for any bracket $\{,\}$ which itself satisfies the Jacobi
identity (see \cite{pjmLB09,pjmCT12}).

Analogous to \eqref{PB}, associated with a bracket of the form of \eqref{DPB}
is a Dirac Poisson operator, which we denote by $\mathbb{J}_\mathbb{D}$.  The relaxation
dynamics is obtained by using a velocity field for advection obtained by
essentially squaring $\mathbb{J}_\mathbb{D}$.  For PV-like dynamics we refer the reader
to \cite{pjmF11}.  In Section 5 we tell this story more explicitly for the contour dynamics that we use in the present work.  Before doing so, we first consider the stability of jets in the $1\frac34$-layer model setting, followed by a discussion of localized steady states.

\section{{\bf3.} Deep and shallow jets: stability}

\subsection{3.A General form}

Equilibrium states in some frame of reference satisfy $\delta\mathfrak{F}/{\delta q}=0$, where 
\bq
\mathfrak{F}[q]=H[q]-\lambda_\mu P_\mu[q]+\calC[q]\,,
\label{VP}
\eq 
with $P_\mu $ defined by \eqref{Pi} being  linear or angular momenta depending on the choice of the function $\phi_\mu(\x)$, which correspond respectively to jets and vortices,   $\calC$ being a Casimir invariant yet to be
chosen, and  $\lambda_\mu$ is a Lagrange multiplier for a chosen  $\phi_\mu$.   We split up $q=Q+q'$,  with the associated $\psi=\Psi+\psi'$, where
\bq
(\nabla^2  -F)\Psi=Q-T \quad \mathrm{and}\quad (\nabla^2  -F)\psi'=q'\,.  
\eq 
Then, if  $Q$  is  an equilibrium solution of \eqref{upper} in a uniformly translating or rotating  frame,  corresponding  to a jet or a vortex, it  solves 
\bq
[\Psi + \la_\mu\ph_\mu, Q]=0\,.
\label{Qequil}
\eq
We these assumptions,  \eqref{VP} gives
\bqy 
\De\mathfrak{F}[q']\equiv \mathfrak{F}[q]- \mathfrak{F}[Q]\!&=&\!\!
H[Q+q']-H[Q] -\lambda_\mu P_\mu [Q+q']+\lambda_\mu  P_\mu [Q] +\int\! d\x \left(C(Q+q')- C(Q)\right)
\nonumber\\ 
\!&=&\!\!- \int\! d\x\!\int\! d\x' \left( q'G(Q-T) + q'Gq'/2 \right)-\lambda_\mu \int \ph_\mu q' + 
\int\! d\x  \left(C(Q+q')-  C(Q)\right) \nonumber\\ 
\!&=&\!\!- \int\! d\x \, q'(\Psi+\lambda_\mu \ph_\mu  ) -  \int\! d\x\!\int\! d\x' \, 
q'Gq'/2+ \int\! d\x \left(C(Q+q')-  C(Q)\right) \nonumber\,.  
\eqy 
Equation \eqref{Qequil} implies  $\Psi + \la_\mu \ph_\mu $ and $Q$ are functionally related -- if we make the choice $C'(Q)=\Psi+\lambda_\mu \ph_\mu$ then 
\bq 
\De\mathfrak{F}[q']= -\half\int\! d\x\!\int\! d\x' q'Gq' + \int\! d\x\!
\big(C(Q+q')-C(Q)-C'(Q)q'\big)\,.
\label{difVP}
\eq 
Thus, the quantity $\De\mathfrak{F}[q']$ can serve as a Lyapunov functional
for stability, by the energy-Casimir method.  (See e.g.\ \cite{pjm98} for
review and early references.  See also \cite{pjmB02,pjmHH14} for detailed
discussion.)

The flow will be stable if $q=Q$ is an extremum; i.e., if
$$
I(q')\equiv C(Q+q')-C(Q)-C'(Q)q' = \int_0^{q'} ds \big(C'(Q+s)-C'(Q)\big)
$$ 
is either positive (except, of course, for $q'=0$) or sufficiently negative
to overcome the first term in \eqref{difVP}, which is positive.

Suppose $C''(q)\ge D_0 \ge 0$; then we have the standard mean value theorem result: 
$$ 
C'(Q+s)-C'(Q) = sC''(Q+s')
$$
with $s'$ between 0 and $s$, and
$$
I(q') \ge D_0\int_0^{q'} ds\;s = {D_0\o 2} q'^2 \ge 0\,, 
$$
so that the flow is stable if $\frac{d}{dQ}(\Psi+\lambda_\mu \ph_\mu) \ge 0$.
This is often referred to as Arnold's first theorem (A-1).

For the $1{\frac34}$ layer model, we can also find cases with $\De\mathfrak{F}[q']$
negative because of the existence of a bound on the energy and PV terms: In
particular,
$$
E[q']\equiv  -  \half\int\! d\x\!\int\! d\x' \, q'G q' < {1\o F} Z[q']
$$
where $Z$ is defined by \eqref{Z}. This follows from
\bqy
Z[q'] &=& \half\int\! d\x\, q'(\nabla^2 \psi'-F\psi')= \half\int\! d\x\, (\del^2 \psi'-F\psi')\del^2\psi'  -\frac{F}{2}\int\! d\x\,q'  \psi' 
\nonumber \\
&=&  \half\int\! d\x \left((\del^2\psi')^2+F |\nabla\psi'|^2 \right)+F E[q']\,,
\label{Z2}
\eqy 
where the first  term of \eqref{Z2} is clearly positive definite.  This  means  $\De\mathfrak{F}[q']$ of \eqref{difVP},  will be negative if we have
sufficiently negative values of $C''$, i.e., if $C''(q)\le D_1\le -F^{-1}$, then
$$
I(q') \le {D_1\o 2} q'^2
~~~{\rm and}~~~
\int\! d\x\, I(q') \le D_1 Z[q'] \le -{1\o F}Z[q'] \,.
$$
Therefore
$$
\De\mathfrak{F}[q']= E[q'] + \int\! d\x\, I(q') \le {1\o F} Z[q'] + \int\! d\x\, I(q') \le {1\o F} Z - {1\o F}Z=0
$$ 
and the flow will also be stable; this is often referred to as Arnold's
second theorem (A-2).

Linearized theory would end up in the same place with $I$ just replaced by the
Taylor-series expansion
$$
\int\! d\x\, I(q) \simeq \half \int\! d\x\, C''(Q) q'^2 \,. 
$$

\subsection{3.B  Vortices and Jets with a linear PV--streamfunction relationship}

A particularly simple choice, solutions of which we call ``linear structures'', has
\bq
Q=\del^2\Psi -F\Psi +T = -{\alpha} [\Psi+\lambda_\mu \phi_\mu ]
\label{linbg}
\eq
and
$$
\mathfrak{F}[q] = -\half\int\! d\x\,\int\! d\x'\,(q-T)G(q-T) 
-\lambda_\mu \int\! d\x\, q\phi_\mu  -{1\o2\alpha}\int\! d\x\,q^2\,.
$$
Again, when we split into the background being a jet or a  vortex, the term linear
in $q'$ clearly vanishes, giving 
\bq
\De\mathfrak{F}[q']=\mathfrak{F}[q]- \mathfrak{F}[Q] = -\half\intint d\x\,d\x'\, q'Gq'
-\frac{1}{2\alpha}\int d\x\, q'^2 =: E[q']-\frac{1}{\alpha} Z[q']\,.
\label{EC}
\eq

The  expression of  \eqref{EC} is  positive definite if $\alpha<0$, 
whence we infer stability.  But the flow will also be stable if
$\De\mathfrak{F}[q']\le 0$ with equality occurring only at $q'=0$.  
We use
$$
(\del^2-F)\psi'=q'
$$
and the  Fourier decomposition of $q'$, denoted $\hat{q}'$,  to obtain 
$$ 
E'-\frac{1}{\alpha} Z'\propto\int d\k\, |\hat
q'|^2\left[\frac{1}{|\k|^2+F}-\frac{1}{\alpha}\right]\,,
$$
which will be negative if  $\alpha < F$.  (Alternatively, the Poincar\'e inequality would yield this result.)
Thus for linear structures we have the following two conditions for stability:
\bq
\alpha<0 \qquad \mathrm{or}\qquad 0<\alpha < F \,.
\label{ECstab}
\eq

For the specific case of  zonal jets moving at speed $c$, we have $\phi=y$ and 
$$
\frac{\p^2}{\p y^2}\Psi -F\Psi +T(y) = -\alpha(\Psi+ c y)
$$
or
\bq
F\psi_2(y) = -\frac{\p^2}{\p y^2}\Psi +F\Psi-\alpha\Psi -(\beta+ c \alpha)y
\label{fpsi}
\eq 
since $T=F\psi_2+\beta y$.  If $\Psi$ has a term linear in $y$ plus a
periodic function, corresponding to a zonal flow that is periodic, $\psi_2$
will have the same character.  This still leaves a great deal of freedom to
choose the structure of the upper layer flows under the presumption that the
deep flows are not well known.  As noted before, $c$ can be viewed as the rate of
translation of the reference frame and will be the rate the vortices are
moving when we look for time-independent eddies atop zonal flows.
For simplicity, however, we will take for the upper layer
\bq
\Psi = \cos(y)-U_1y
\label{PsiU}
\eq (setting both the length and velocity scales for nondimensionalization
based on the upper layer jets).  Then substituting \eqref{PsiU} into
\eqref{fpsi} gives for the lower layer 
\bq
\psi_2 = D\cos(y)-U_2 y
\label{psiL}
\eq
with
\bq
D = 1+\frac{1-\alpha}{F}
\qquad\mathrm{and}\qquad
U_2 = U_1+\frac{\beta+\alpha c-\alpha U_1}{F}\,.
\label{linalpha}
\eq

\begin{figure}[htb]
\centering
\includegraphics[width=9 cm]{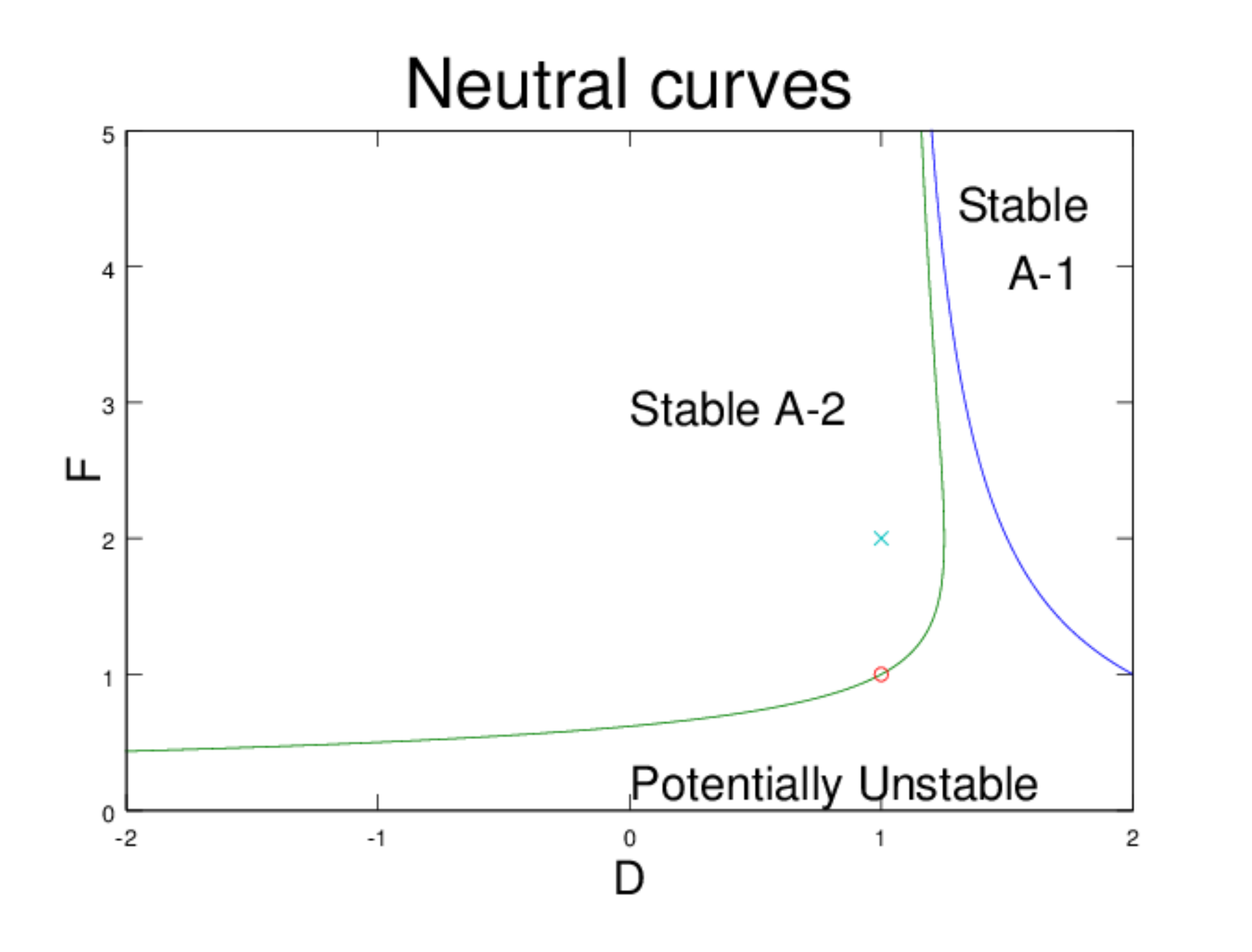}\\
\caption{Regimes that are stable, based on $\mathfrak{F}[q]$ being positive
definite (marked A-1) or negative definite (marked A-2).  Most of the
derivations and simulations take the deep and shallow jets to match ($D=1$).
The symbol $\times$ indicates the standard values $D=1, F=2$, well above the
transition to stability at $F=1$, marked by $\circ$.}
\label{stabB}
\end{figure}

Figure \ref{stabB} shows the regions of stability for this solution,
$\alpha<0$ or $0<\alpha<F$, in terms of $F$ and the ratio $D$ of the deep to
the upper layer jet strength.
The best and simplest guess from available information from the Galileo probe
\cite{dowling89} is that $D=1$ which occurs for $\alpha=1$.  In that case, the
flow will be stable if $F>1$ meaning the length scale of the jets is larger
than the deformation radius $F^{-1/2}$.

Intimately related to the energy-Casimir method is a Rayleigh criterion (see Refs.~\cite{pjmB02} and \cite{pjmH14} for general discussion);
taking $C'(Q) = \Psi+c y$,  A-1 is obtained if $C''(Q)>0$. However,
\[
C''(Q)\,Q_y  = -(U(y)-c)
\qquad\mathrm{and}\qquad 
U(y)=-\frac{d\Psi}{d y}
\]
so that we can make the flow stable if $Q_y$ does not change sign. 
Thus sufficiently strong $\beta$, i.e., greater than one in the case $D=1$,  will
also stabilize the flow.

Finally, we note that the $1\frac{1}{2}$ layer model just has $T=\beta y$ and
$\psi_2=0$ so that (\ref{fpsi}) in the sinusoidal case implies $\alpha=1+F$
and (A.2) cannot hold.  When $\beta<1$, the flow is indeed unstable, and the
jets break up.

\section{{\bf4.} Steady state vortices}

\subsection{{\it 4.A.} Localized vortex in jets}

The  $1{\frac34}$-layer model written relative to the stationary jets in a frame moving at speed $c$ takes the form, 
\bq
\ppt q' + [\Psi+cy+\psi',Q+q']=0 \qquad \mathrm{with}\qquad q'=(\del^2-F)\psi'\,.
\label{qpDyn}
\eq
We seek  steady states of this system  that satisfy \eqref{qpDyn} with $\p q'/\p t$ =0\,. 
In particular, we are interested in localized vorticity anomalies,
which from  \eqref{qpDyn}  must satisfy
\bq
Q+q' = \Qf(\Psi+cy+\psi')
\label{eq1}
\eq 
for an arbitrary function $\Qf$.  If the vortex is decaying in all
directions,  the vanishing of $q'$ and $\psi'$ far from the vortex center  implies 
\bq
Q=\Qf(\Psi+cy)
\label{QPSI}
\eq
and therefore from \eqref{eq1} 
\bq
q' = \Qf(\Psi+cy+\psi')-\Qf(\Psi+cy)
\label{eq1q}
\eq
outside closed streamlines.  Far from the vortex center,
$$
q'\rightarrow \Qf'(\Psi+cy) \psi'
\quad {\rm or}\quad
\del^2\psi'-F\psi'  = {1\o C''(Q)}\psi' \,. 
$$
The conditions for stability $C''>0$ or $C''< -1/F$ ensure that $\del^2\psi'$
is everywhere related to $\psi'$ by a positive coefficient.  So we don't have
far field wavelike behavior and can expect to find isolated vortex solutions.

\subsection{4.B Linear case}

For linear  background flow, where $C''= -1/\alpha$ and
$$
\del^2\psi' = (F-\alpha)\psi'\,,
$$
one has decaying modified Bessel function solutions for $F>\alpha$.  The conditions
for stability are exactly those that permit isolated vortices to be embedded
in the flow.  The case of sinusoidal jets just has $\alpha=1$.

For comparison, in the $1\frac{1}{2}$ system, we would have 
$$
(\del^2-F)\psi'=-(1+F)\psi'\,,
$$
which cannot  have isolated solutions in all directions.

\subsection{{\it 4.C.} Integral conditions}

From the $x$-moment of the equations for the PV anomalies
\bqy
\ppt\int\! d\x\, xq' &=& -\int\! d\x \left( x[\Psi+cy,q']+x[\psi',Q]+x[\psi',q']\right)\nonumber \\
&=& -\int\! d\x \left(q'[x,\Psi+cy] -\psi'[x,Q]+q'[x,\psi']\right)\nonumber \\
&=& -\int\! d\x \left( q'(c-U)-\psi' Q_y +q'\psi'_y\right) \nonumber\\
&=& -\int\! d\x \left(\left(c-U\right)\left(q'-\Qf'\psi'\right)\right) \nonumber\\
&=& -\int\! d\x \left( (c-U)\left(q'-\frac{1}{C''(Q)}\psi'\right)\right)\,,
\nonumber
\eqy
where in proceeding from the third to the fourth equality  \eqref{QPSI} was used and the $q'\psi'_y$-term vanishes by integration by parts. 
If we define
\bq
\tq := q'-\frac{1}{C''(Q)}\psi' = \del^2\psi' - \left(F+\frac{1}{C''(Q)})\right)\psi' 
\label{tildeq}
\eq
(which goes to zero in the far field), our moment equation tells us
$$
\ppt\int\! d\x\,  xq' =\int\! d\x\,  U\tq - c\int\! d\x\, \tq\,;
$$
for steady propagation this leads to 
\bq
c=\frac{\int\! d\x \, U\tq}{\int\! d\x\,  \tq}\,. 
\label{ssc}
\eq
The constraint \eqref{ssc} places on the speed has to be considered along with the
constraints implied by (\ref{fpsi}).  In particular, for the linear,
sinusoidal case, (\ref{fpsi}) implies $\alpha=-1/C''(Q)=1$ and $c=-\beta$.  A
small vortex must then be centered where $U=c=-\beta$.  Note that the zero in
the denominator of
$$
\Qf' = \frac{\beta-U_{yy}}{c-U} = \frac{\beta+U}{c-U}
$$
matches with the zero in the numerator so that $\Qf'=-1$ everywhere.

\subsection{4.D Linear jets}

Inserting $Q=-\alpha(\Psi+ c y)$ into  \eqref{qpDyn} gives
\bqy
\ppt q' &=& -[\Psi+cy+\psi',-\alpha(\Psi+cy) + q']
\nonumber\\
&=&-[\Psi+cy+\psi',-\alpha(\Psi+cy+\psi') + q'+\alpha\psi']
\nonumber\\
&=&-[\Psi+cy+\psi',q'+\al \psi'] 
\nonumber\\
&=&- [\Psi+cy+\psi',\tq]\,,
\label{qpDyn2}
\eqy
where the  final form that uses \eqref{tildeq} is convenient for  seeking steady states, as we shall see in Section 4.D. We will be looking for contour dynamics type solutions with
$\tq = q_0$ within an area $\cala$ and zero outside.  We then want to solve
$$
[\del^2 -(F-\alpha)]\psi' = q_0
~~~{\rm for}~~~ \x\in\cala\,. 
$$
The integral condition \eqref{ssc} gives
\bq
c=U_1 +\frac{1}{\cala}\int_\cala d\x\, \sin(y)\,,
\label{paIC}
\eq
while the far-field condition, from eliminating $\alpha$ from (\ref{linalpha}),  has
\bq
c=U_1-\frac{\beta+F(U_1-U_2)}{1+F-FD}\,.
\label{full}
\eq
Given the parameters of the background flow, these two expressions for $c$  imply that the
north-south location of the vortex is determined, although it depends on the
precise shape.

For the standard case, $c=-\beta$ and $\int_\cala \sin(y)=-\beta$. For a
small vortex, $\sin(y_c) = -\beta$. When finding steady states, we shall
choose the centriod of the vortex, let the algorithm calculate $c$ which will
satisfy (\ref{ssc}) and then use that to determine the correct $\beta$ value.
We can then use the $\beta(y_c)$ values to find where the vortex will reside
given instead the value of $\beta$.

\subsection{{\it4.C.} Modified dynamics}

From Section 4.D the problem of a vortex with a uniform PV anomaly under the linear jet assumption
amounts to the following choice for $\Qf(Z)$:
$$
\Qf(Z)=-\alpha Z+q_0\chi_\cala(\x)\,,
$$ 
with the characteristic function $\chi_\cala(\x)=1$ when $\x$ is in the
patch area $\cala$ and zero when it is outside.  The boundary of the patch
must be a contour of constant $Z$ and the equation to solve is 
$$
\Big[\del^2-(F-\alpha)\Big]\psi'= q_0\chi_\cala(\x)
$$
with
$$
\Psi+cy +\psi' = {\rm const.~~on~~}\p \cala\,.
$$ 
We will use contour dynamics to evaluate $\u'$ given the boundary shape and
the Dirac-bracket synthetic annealing of \cite{pjmF11} for a modified dynamical system, adapted for contour
dynamics in Section 5, to find the shape.

To put \eqref{qpDyn2} into  a form where the DBSA tools can be applied, we
again write
$\tq=q'+\alpha\psi'$ giving the dynamical equation
\bq
\ppt(\tq-\alpha\psi') + [\Psi+cy+\psi',\tq] = 0
\qquad\mathrm{with}\qquad 
\tq=(\del^2-F+\alpha)\psi'\,.
\label{tDyn}
\eq
However, if we are only interested in the steady states, we can just as well consider 
the modified dynamics of 
\bq
\ppt \tq + [\Psi+cy+\psi',\tq] = 0\,,
\label{mDyn}
\eq because \eqref{tDyn} and \eqref{mDyn} possess the same steady states: if
we construct a simulating annealing dynamics that relaxes to steady states of
\eqref{mDyn}, then we obtain steady states of \eqref{tDyn}.  {However, it should be borne in mind that unlike \eqref{upper}, which preserves $q$ on particles,  \eqref{mDyn} preserves $\tilde{q}$ on particles, and our DBSA algorithm will do this as well.} For the case of  
interest here, where $\tq$ is a piecewise constant patch, in the modified
dynamics the anomaly within the patch and the area of the patch are conserved, 
whereas that is not the case for the original equation \eqref{tDyn}.
Therefore, the modified dynamics can be treated by the methods of Hamiltonian
contour dynamics that we describe next.

\section{{\bf5.} Hamiltonian  contour  dynamics and synthetic annealing}

\subsection{{\it 5.A.} Hamiltonian Structure of Contour Dynamics}
\label{ssec:hamCD}

The equations of contour dynamics \cite{roberts1,roberts2} are an example of a
reduction of the two-dimensional Euler fluid equations.  Consequently, contour
dynamics inherits the Hamiltonian structure of vortex dynamics (see
e.g.\ \cite{pjm82,pjm98}), and can in fact be derived therefrom.  The
reduction is based on initial conditions where the dynamical variable is
constant in a region bounded by a contour.  Then for transport equations like
that for two-dimensional vorticity it is known that this structure is
preserved in time, with the dynamics restricted to be that of the moving
bounding contour.  The Hamiltonian structure of interest here is one with a
noncanonical Poisson bracket, like that of Section 2, that does not require
the contour bound a star-shaped region, i.e.,  the contour need not have a parameterization as a graph of an angle.  Indeed the bounding contour is any plane curve (or curves) with an arbitrary parameterization.
Here we will first describe the situation for the two-dimensional Euler fluid
equations, where the scalar vorticity $\omega(\x,t)=\nabla^2\varphi$ is constant inside the contour, before showing how
to apply this to the case of interest here.

The reduction to contour dynamics proceeds by replacing  $\om$ by a plane curve that bounds a vortex patch
or patches $\vX(\sigma)=(X(\sigma),Y(\sigma))$.  Here the curve parameter
$\sigma$ is not chosen to be arc length because arc length is not conserved by
the dynamics of interest.

Because plane curves are geometrical objects, their Hamiltonian theory should
be based on parameterization invariant functionals, i.e. functionals of the
form
$$
A[X,Y]=\oint d\si\, {\cal A} (X,Y,X_\si,Y_\si, Y_{\si\si},
 X_{\si\si},\dots)\,,
$$ 
where $X_{\sigma} :=\p X/\p \sigma$, etc.  and ${\cal A}$ has an Euler
homogeneity property making such functionals invariant under
reparameterization $\si(\si')$ say.  A consequence of parameterization
invariance is the Bianchi-like identity that ties together functional
derivatives, 
\bq 
{\de A\o\de X(\si)}\;X_\si + {\de A\o\de Y(\si)}\;Y_\si\equiv 0\,.
\label{constr}
\eq 
The constraint of \eqref{constr} can be compactly written as
$\hat{\tau}\cdot {\de A}/{\de \bfX}=0$, where
$\hat{\tau}=(X_\si,Y_\si)/\ntan=\bfX_\si/\ntan$, with $\ntan^2=X_\si^2 +
Y_\si^2$, is the unit vector tangent to the contour and $\de A/\de \bfX:= (\de
A/\de X, \de {A}/\de Y)$.  This is a result that follows from E.~Noether's
second theorem \cite{noether18}.

The noncanonical Poisson bracket for the contours is given, in its most
symmetrical form, by 
\bq 
\{A,B\} = \bigointsss d\si\, \left[Y_\si {\delta
 A\o\delta X} - X_\si {\delta A\o\delta Y} \o X_\si^2 + Y_\si^2\right]\;
{\p\o \p\si} \; \left[Y_\si {\delta B\o\delta X} - X_\si {\delta B\o\delta Y}
\o X_\si^2 + Y_\si^2\right]\,,
\label{cd.pb}
\eq 
where we assume closed contours, although generalizations are possible. 
Observe that if the two functionals $A$ and $B$ are parameterization
invariant, then so is their bracket $\{A,B\}$.  The bracket of \eqref{cd.pb}
can be rewritten as
\[
\{A,B\} =  
\oint d\si\,\, 
\frac{\de A}{\de \bfX}\cdot
\mathbb{J}_{\mathrm{CD}}
\cdot \frac{\de B}{\de \bfX}\,, 
\]
where the noncanonical Poisson operator $\mathbb{J}$ for this bracket is the
following skew-symmetric matrix operator:
\[
\mathbb{J}_{\mathrm{CD}}= 
\left( 
\begin{array}{cc}
\frac{Y_\si}{\ntan^2}\frac{\p }{\p \si} \frac{Y_\si}{\ntan^2} & -
\frac{Y_\si}{\ntan^2}\frac{\p }{\p \si} \frac{X_\si}{\ntan^2}
\\ -\frac{X_\si}{\ntan^2}\frac{\p }{\p \si} \frac{Y_\si}{\ntan^2} &
\frac{X_\si}{\ntan^2}\frac{\p }{\p \si} \frac{X_\si}{\ntan^2}
\end{array}
\right)\,.
\]

The dynamical equations for the contour are generated by inserting the
following compact form for the Hamiltonian into the bracket of \eqref{cd.pb}:
\bq H = \oint d\si \oint d\si'\phi~ \hat{\tau}\cdot \hat{\tau}'\,,
\label{Cham}
\eq where $\hat{\tau}$ and $\hat{\tau}'$ are the unit vectors tangent to the
contour, with $\tau'$ being that for the contour parameterized by $\si'$, and
$\phi(\rho)$ satisfies ${\nabla'}^2 \phi(\rho) = G(\rho)$, where
$\rho=|\bfx-\vxp|$ and $G$ is the two-dimensional Green's function. Note, in
\eqref{Cham} the argument of $\phi$ is $|\bfX-\bfX'|$.  This Hamiltonian can
be obtained from that for the two-dimensional Euler equation by restricting to
patch-like initial conditions and manipulating; furthermore, it can be shown
to be parametrization invariant in accordance with our theory.

Upon insertion of \eqref{Cham} into \eqref{cd.pb} we obtain the contour
dynamics equations of motion, \bqy \dot\bfX&=&\{\bfX,H\}=
\frac{\hat{n}}{\ntan} \frac{\p}{\p \si} \frac{\hat{n}}{\ntan}\cdot
\frac{\delta H}{\de \bfX} \nonumber\\ &=&
\left(\frac{\ys}{\rts},-\frac{\xs}{\rts}\right) \frac{\p}{\p \si}
\varphi(\bfX,t)= (u(\bfX,t),v(\bfX,t))\,,
\label{VF}
\eqy
where $\hat{n}= (Y_\si,-X_\si)/\ntan$ is the unit outward normal. 

The area of a patch is evidently given as follows: 
\bq
\Ga:=\int_D d\bfx=\half\int d\bfx \, \nabla\cdot \bfx= \half\oint d\si\, 
(XY_\si - Y X_\si)\, ,
\label{Ga}
\eq yielding finally a contour functional that is easily shown to be a Casimir
invariant for the bracket \eqref{cd.pb}, i.e.\ $\{\Ga,B\}\equiv 0$ for all
functionals $B$.  Thus, a class of Hamiltonian field theories on closed curves
that preserve area is defined by \eqref{cd.pb}, for {\it any} Hamiltonian
functional.  Note, however, although area is preserved perimeter is not.  Also
note, $\Ga$, like all functionals admissible in this theory,  is parametrization
invariant.

Starting from an expression for the angular momentum $L$ (physically minus the angular momentum) of \eqref{Pi} one
can reduce to obtain its contour dynamics form \bqy L&=&\int_{D}d\bfx\, (x^{2}
+ y^{2}) = \frac{1}{4} \int_{D}d\bfx\, \nabla\cdot\left(\bfx \,(x^{2} + y^{2})
\right) \nonumber\\ &=& \frac{1}{4} \oi (X^2+Y^2)\left(X\ys -Y\xs\right)\,,
\label{cdAngMom}
\eqy which is a dynamical invariant following from Noether's first theorem,
i.e., an invariant not due to bracket degeneracy but tied to the Hamiltonian
of interest \eqref{Cham} via $\{L,H\}=0$.  Again observe $L$ is
parametrization invariant, as expected on physical grounds.

\subsection{5.B. Dirac Brackets and Simulated Annealing}

Now we use the development of Section 5.A to construct a Dirac bracket analogous
to that of \eqref{DPB} of Section 2.C, in order to apply our DBSA method.  This is the contour dynamics version of the
procedure in \cite{pjmF11} with the Dirac bracket constructed using the
contour dynamics bracket of \eqref{cd.pb}.  This will yield a system of the
form \bq \frac{d X^i}{d t}= \et_H\, \mathbb{J}^{ij}_{\mathrm{CDD}}\, \frac{\de
\mathfrak{F}}{\de X^j} + \et_{SA}\, \mathbb{J}^{ij}_{\mathrm{CDD}}\,
\mathbb{G}_{jk}\, \mathbb{J}^{kj}_{\mathrm{CDD}}\, \frac{\de \mathfrak{F}}{\de
X^j}\,,
\label{cddbsa}
\eq where we have set $X^1=X$ and $X^2= Y$ and used repeated index notation
over $j,k=1,2$, $\et_H$ and $\et_{SA}$ are numbers that weight the
contributions of each of the terms of \eqref{cddbsa} to the dynamics,
$\mathfrak{F}[\bfX]$ is a functional analogous to that of \eqref{VP} but now
for contour dynamics, $\mathbb{G}$ is a symmetric smoothing metric that we are
free to choose and, most importantly, $\mathbb{J}_{\mathrm{CDD}}$ is the
Poisson operator that arises from \eqref{DPB} upon using \eqref{cd.pb} with
constraints $C_{1,2}$, constraints that we will choose explicitly below.  We found in
\cite{pjmF11} that in order to obtain a rich class of steady states it was
necessary to use the Dirac constraints $C_{1,2}$ chosen judiciously for the
desired state.  Ordinary SA corresponds to the case where $\mathbb{J}_{\mathrm{CD}}$ is used in \eqref{cddbsa}, while DBSA uses $\mathbb{J}_{\mathrm{CDD}}$ that enforces the Dirac constraints.

Relaxation proceeds under the dynamics of \eqref{cddbsa} in a manner analogous
to the H-theorem relaxation of the Boltzmann equation to thermal equilibrium.
The functional $\mathfrak{F}$ generates relaxation dynamics to
$\de\mathfrak{F}=0$ because $d\mathfrak{F}/dt \geq 0$, which follows from
\eqref{cddbsa}.  In Section 5.C, we will demonstrate how this works in the
simpler context of plain contour dynamics, before using this simulated
annealing technique to calculate Jovian vortices in Section 5.D.

\subsection{{\it 5.C.} The Kirchhoff ellipse with shear}

As a first example we calculate the Kirchhoff ellipse, the well-known exact 
solution of the two-dimensional incompressible Euler equation.  Because there
are many steady states in rotating frames, e.g., the V-states of rigidly
rotating vortex patches with $m$-fold symmetry \cite{deem78}, something more
is need to select out the Kirchhoff ellipse.  Extremization of
$\de\mathfrak{F}=\delta(H-\Omega L)=0$, where $L$ is given by \eqref{cdAngMom}, an expression for
the angular momentum,  and $\Om$ a constant, is insufficient -- it need not
preserve $L$ and simply goes to a circle.  For this reason we employ Dirac constraints in our DBSA algorithm.
To this end we choose $L$, already an invariant, to play a dual role as one of
our Dirac constraints.  The other Dirac constraint is chosen as in
\cite{pjmF11} to be the $xy$-moment that enforces 2-fold symmetry.  For
contour dynamics this is \bq K=2\int_{D}d\x\, xy =\frac1{2}\oi \, XY
\left(X\ys -Y\xs\right)\,.
\label{K}
\eq
These are viable constraints because $\{L,K\}\neq 0$, a Dirac bracket requirement that ensures the denominator of   \eqref{DPB} does not vanish. 

Because the Dirac bracket is complicated we introduce the following shorthand notation: 
\bq \frac{\delta F}{\delta \om}:= \fq 
\label{short}
\eq
and obtain by direct calculation
\bq
\frac{\delta L}{\delta \om}= \rt 
\qquad \mathrm{and}\qquad
\frac{\delta K}{\delta \om}= 2XY\,,
\label{first}
\eq
which give 
$$
\{L,K\}=\oi\left(\rt\right) \frac{\p}{\p\si} \left(2XY\right)\,.
$$ 
Similarly, we obtain 
\begin{align}
\{F,L\}&=\oi 
\frac{\delta F}{\delta q} \frac{\p}{\p\si} (\rt) 
&
\{G,K\}&=\oi 
\frac{\delta G}{\delta q} \frac{\p}{\p\si} (2XY) 
\\ 
\{X,L\}&=\frac{\ys}{\rts}\frac{\p}{\p\si} (\rt)
&\{Y,L\}&=-\frac{\xs}{\rts}\frac{\p}{\p\si} (\rt)
\\
\{X,K\}&=\frac{\ys}{\rts}\frac{\p}{\p\si} (2XY) 
&\{Y,K\}&=-\frac{\xs}{\rts}\frac{\p}{\p\si} (2XY)\,,
\end{align}
which when inserted into
\eqref{DPB} with $\mathfrak{F}$ give the following vector field analagous to
\eqref{VF}:
\begin{eqnarray}
\left(u_{SA},  v_{SA}\right)&=& \pm\left(\frac{\ys}{\rts}, 
-\frac{\xs}{\rts}\right)
\nonumber\\
&\times&\Bigg[ \Phi+\frac{\p}{\p\si} (\rt)
\frac{\oi \left(2XY\right) \Phi}
{\oi\left(\rt\right) \frac{\p}{\p\si} \left(2XY\right)}
\nonumber\\
&&\hspace{3 cm} 
-\frac{\p}{\p\si} (2XY)
\frac{\oi \left(\rt\right) \Phi}
{\oi\left(\rt\right) \frac{\p}{\p\si} \left(2XY\right)}\Bigg]\,,
\label{SAvf}
\end{eqnarray}
with
$$
\Phi := \frac{\ys u}{\rts}-\frac{\xs v}{\rts}\,.
$$ Here the parameters $\et_H$ and $\et_{SA}$ have been set to zero and unity,
respectively.

\begin{figure}[htb]
\centering
\subfigure[{\footnotesize \  }]{\includegraphics[width=0.49\textwidth]{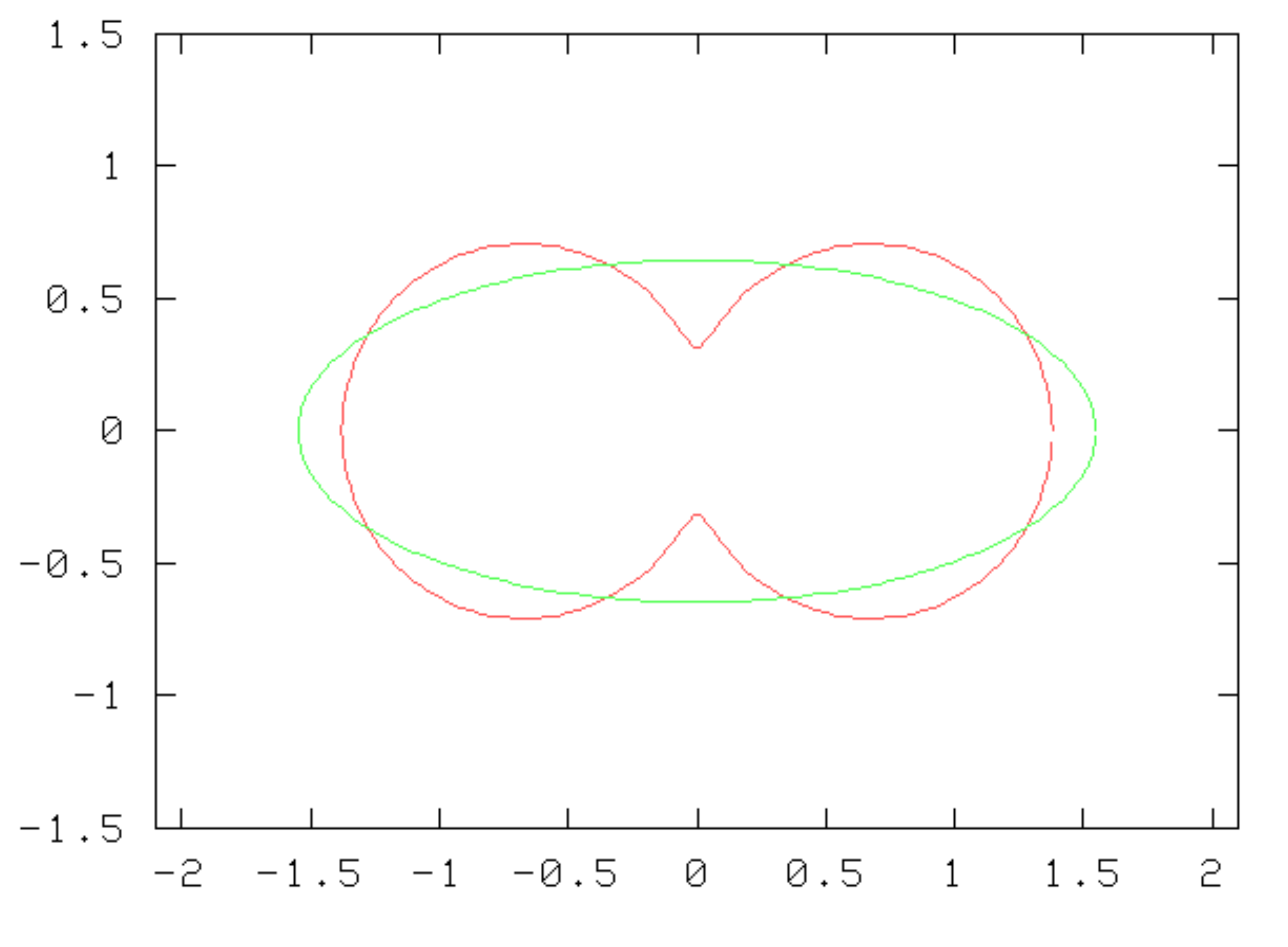}
\label{fig:kirch}
}
\subfigure[{\footnotesize \  }]{\includegraphics[width=0.47\textwidth]{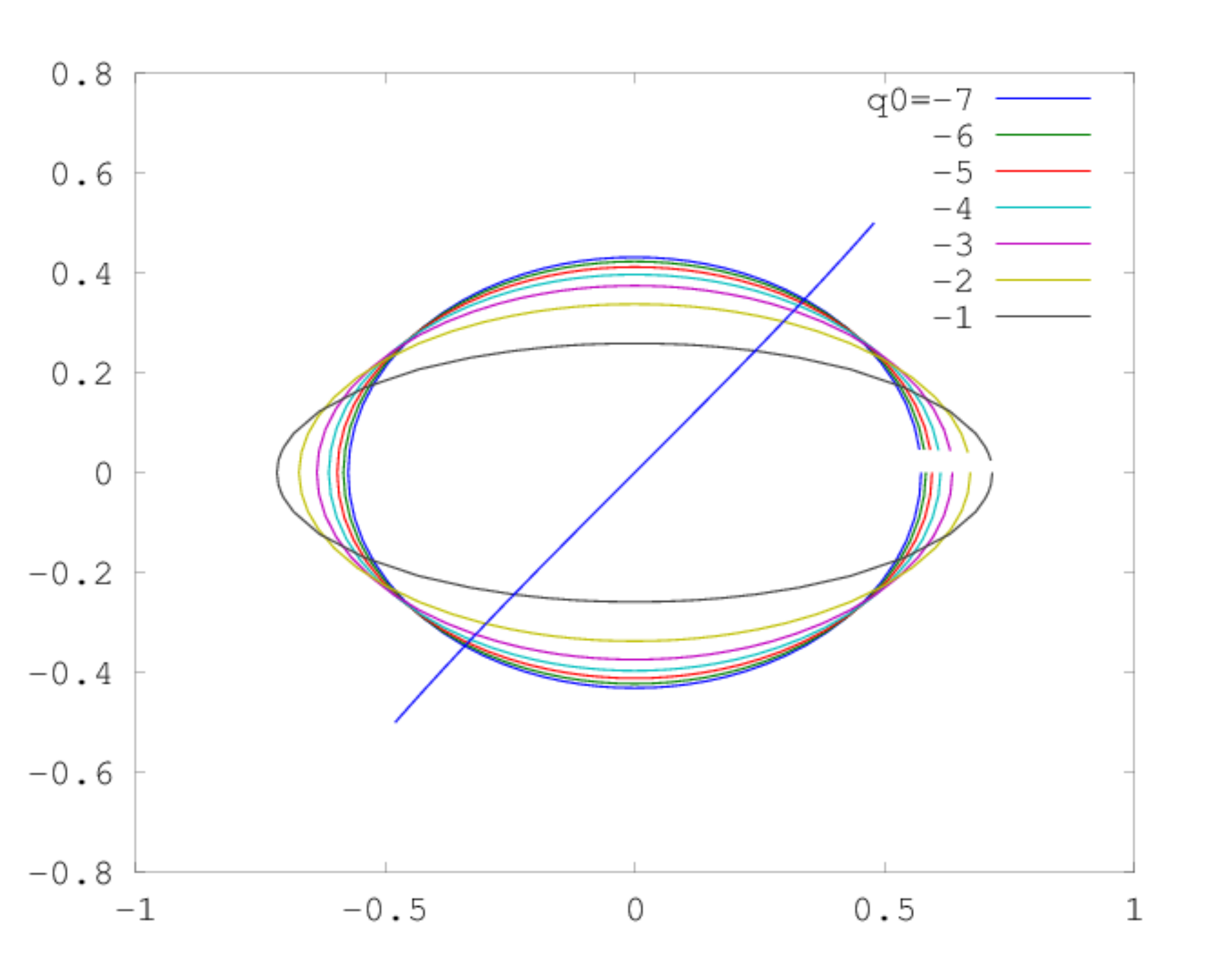}
\label{fig:kida}
}
\caption[sm] {DBSA for contour dynamics with the two constraints $L$ and $K$
of \eqref{cdAngMom} and \eqref{K}, respectively.  (a) Relaxation to the
elliptical V-state with 2-fold symmetry (Kirchhoff ellipse). (b) Relaxation
to an elliptical state in the presence of shear (Kida ellipse) for various anomalies $q_0$.  }
\label{fig:CD}
\end{figure}

Figure \ref{fig:CD} depicts two results for this case.
Figure \ref{fig:kirch} shows an initial ``dog bone" state relaxing to the
Kirckhoff ellipse.  In Figure \ref{fig:kida} a background linear shear flow has
been added to the Hamiltonian functional (see \cite{pjmMF97}) and the
relaxation to a variety of Moore and Saffman \cite{moore} ellipses -- the steady version
of Kida ellipses \cite{kida} -- are obtained.

\subsection{5.D. Application to Jovian vortices}

Let us now describe how the formalism can be used to calculate Jovian vortices.  For the modified dynamics of (\ref{mDyn}),  vortex states will depend on the strength of the
anomaly, $q_0$, the initial radius, $r_0$ with $\cala=\pi r_0^2$, and the initial
center latitude $y_0$. We use $\Psi=\cos(y)$ and let the procedure determine
what the changes in $c$ and $U_1$ need to be in order to have a steady
equilibrium.   We will use the linear momenta as Dirac constraints, which to within a sign are
$C_1=\int x\tq$ and $C_2=\int y\tq$.  The DBSA routine then makes
\bq
[\cos(y)+\psi'+a_1 x+a_2 y,\tq]=0\,.
\label{JV}
\eq
For this case the integral conditions (cf.\ Section 4.C) are obtained by multiplying  \eqref{JV} by $y$ and $x$ and integrating, yielding respectively 
\begin{eqnarray*}
\int\! d\x \, y [a_1 x+a_2 y,\tq]&=&  - a_1\int\! d\x \,  \tq= \int\! d\x \, y[\cos(y)+\psi',\tq] =- \int\! d\x \,\tq\ppx\psi'=0\\
\int\! d\x \, x [a_1 x+a_2 y,\tq]&=&a_2 \int\! d\x \,\tq = \int\! d\x \, x[\cos(y)+\psi',\tq] =\int\! d\x \,\left( \sin(y)\tq + 
\tq\ppy\psi' \right)\\
&=&\int\! d\x \,\sin(y)\tq\,.
\end{eqnarray*}
Consistent with symmetry, $a_1$ will be zero; comparing
the second condition for a vortex patch of area $\cala$ to the integral constraint of \eqref{paIC} of Section 4.D  gives
$$
a_2 =c-U_1\,.
$$ 
To match with the expressions from the full model, according to \eqref{linalpha} this would need to be 
$a_2=-[\beta+F(U_1-U_2)]/\alpha$ (or $-\beta$ in the linear jet case with $U_1=U_2$ and $\al=1$), but that will
generally not be the case.  The DBSA algorithm will give $c(q_0,\cala,y_0)$;
we can adjust $y_0$ so that this agrees with the value stipulated by the
far-field requirement.  We shall simply look at the inverse problem: examining
$\beta(q_0,\cala,y_0)$.

\section{6. Results}

\subsection{6.A. CD--DBSA}

\begin{figure}[htb]
\centering
\subfigure[{\footnotesize \  }]{\includegraphics[width=0.48\textwidth]{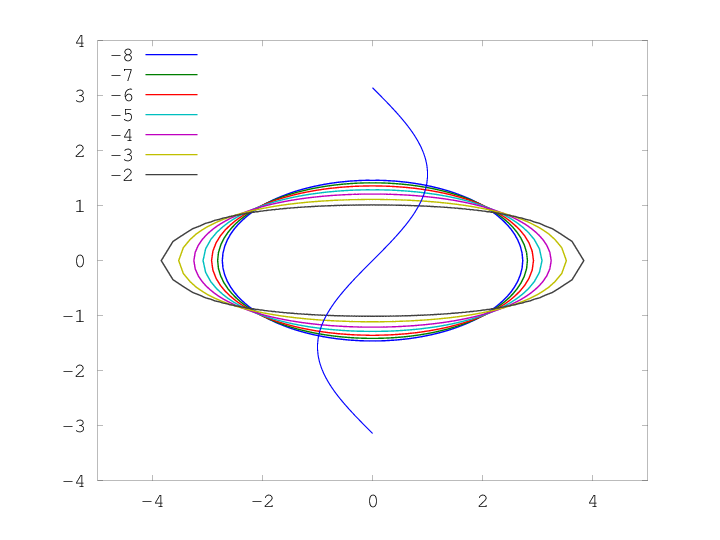}
\label{fig:yzeroA}}
\subfigure[{\footnotesize \  }]{\includegraphics[width=0.48\textwidth]{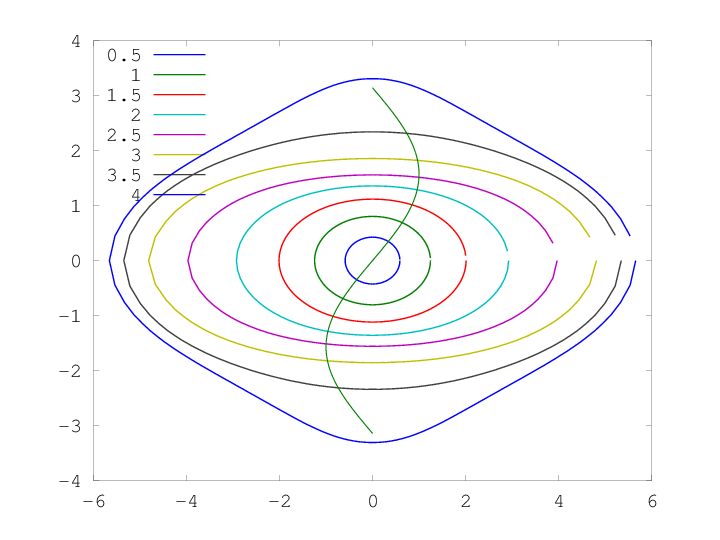}
\label{fig:yzeroB}}
\caption[sm] {DBSA with the two  $x$ and $y$ moments taken as Dirac constraints.  (a) Vortex
shape as a function of $q_0$ from $-8$ to $-2$.  The jet vorticity is $-1$ at the centerline and the area is fixed at $4\pi^2$. (b) Vortex shape as a function of area defined by the  initial radius $r_0$ from $0.5$ to $4$. The PV anomaly is $q_0= -6$. }
\label{fig:yzero}
\end{figure}

We first concentrate on the case in which the background flow does not vary with
depth,  $U_1=U_2=0,~D=1$ so that $\alpha=1$ (cf.\ \eqref{PsiU}, \eqref{psiL}, and \eqref{linalpha}).  When $\beta=0$, the vortex center  resides at  $y=0$  where the background flow changes sign, and it is symmetric
both east-west and north-south as depicted in Figure \ref{fig:yzero}.  Observe in Figure \ref{fig:yzeroA} that it  elongates as $|q_0|$ decreases.
In Figure \ref{fig:yzeroB} where the size increases,  the vortex becomes visibly different from an exact
ellipse, since it feels the changes in the shear.
It is known that the  Moore and Saffman \cite{moore} ellipse in uniform shear (the steady
version of Kida's solution \cite{kida}) has a smaller radius of curvature at
the north and south for weaker shears relative to $q_0$; similarly, the solutions of Figure \ref{fig:yzeroB} 
have more curvature where the shear is small.

Although the  solutions of Figure \ref{fig:yzero} were obtained using the full DBSA algorithm with Dirac constraints $C_{1,2}$ being the $x$ and $y$ moments, they could equally well be found by just applying SA;  unlike the Kirchhoff ellipse \cite{kirchhoff} above which would, in the absence of the Dirac constraints,
become circular. Essentially, the background shear locks in the position and
orientation, with the  amplitudes $a_i$ remaining zero throughout because of the
symmetry.  However, when we consider $\beta \ne 0$ this is no longer the case and
the constraints become critical.

\begin{figure}[htb]
\centering \subfigure[{\footnotesize
 \ }]{\includegraphics[width=0.45\textwidth]{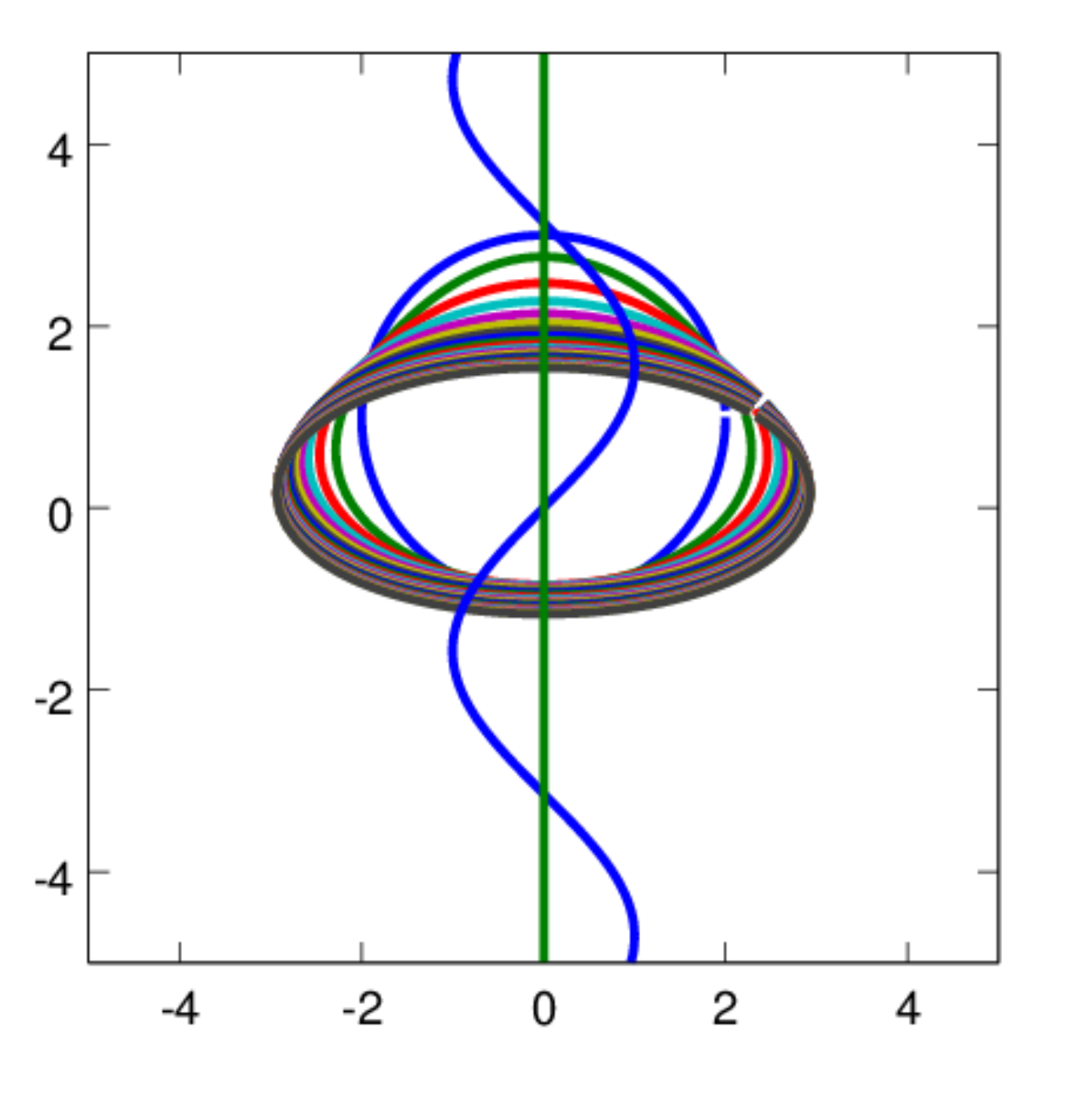}
\label{fig:beta1A}
}
\subfigure[{\footnotesize \  }]{\includegraphics[width=0.45\textwidth]{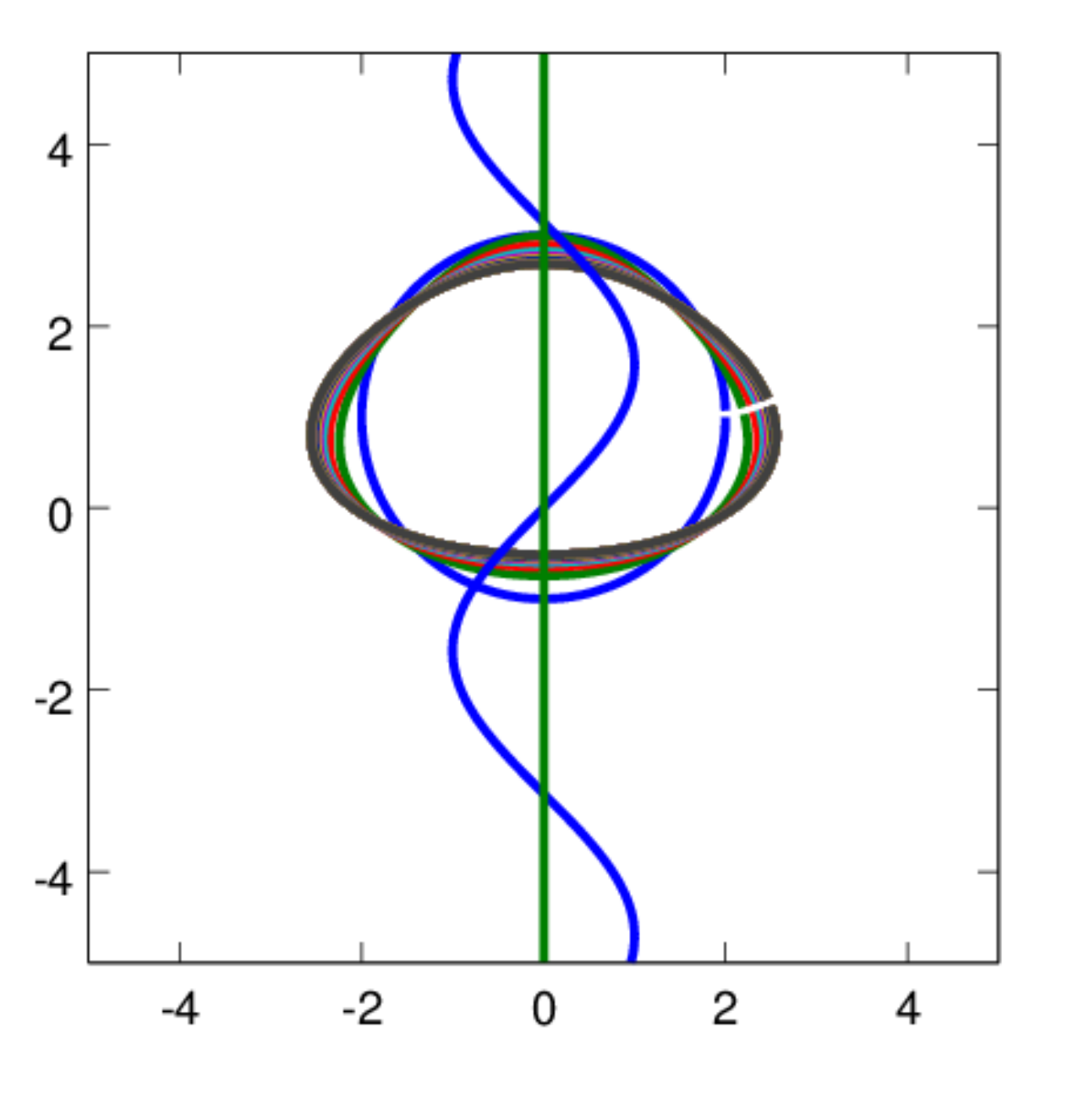}
\label{fig:beta1B}
}
\caption[sm] {Depiction of the evolution from a circular patch with (a) SA
vs.\ (b) DBSA. Initially the patch is off axis (the y value being where the background flow $U$ is
antisymmetrical).  Whereas SA moves it towards the axis, DBSA tries to find
the distorted shape at that center.  
}
\label{fig:beta1}
\end{figure}

\begin{figure}[htp]
\centering
\includegraphics[width=9 cm]{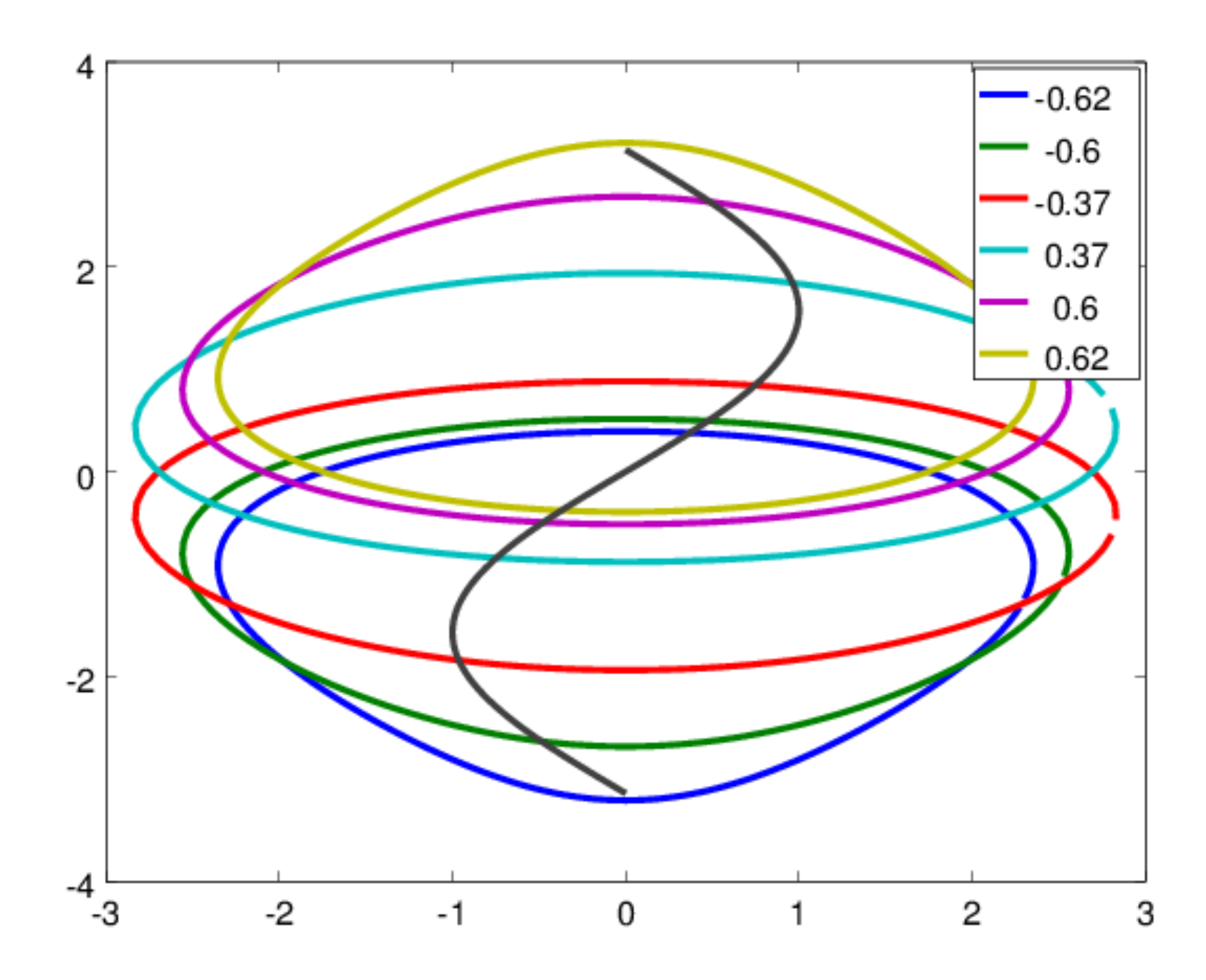}\\
\caption{Vortex shape as a function of $\beta$, with  $q_0$ and the area  fixed.
The vortex is initially centered at $y_0=-1.25,-1,-0.5,0.5,1,1.25$; labels
give the value of $\beta$ (to be compared with $max\left|U''(y)\right|=1$) for which this would be the equilibrium
solution.}
\label{fig:beta2}
\end{figure}

For the  $\beta \ne 0$ case, as mentioned in Section 5.D, we solve this problem in
reverse by starting with a vortex centered at $y=y_0$.  Figure \ref{fig:beta1A} shows that the standard synthetic annealing process moves this down to the $y=0$ axis of symmetry and then reverts to the solution in
Figure \ref{fig:yzero}.  In contrast, the constrained solution, Figure \ref{fig:beta1B}, remains centered at $y_0$ in the sense that the integral
$\int\!d\x\,qy$ is conserved; this is because that is built into the Dirac bracket
as $C_2$.  When the solution settles, we now have a finite value of $a_2$,
which depends on the offset $y_0$ and which then determines the compatible
value of $\beta=-\alpha a_2$. Figure \ref{fig:beta2} shows a set of shapes and
the corresponding $\beta$ values.  As $\beta$ increases, the vortex resides
further off the axis and has more of the $m=3$ triangular mode.

Jupiter's Great Red Spot is, of course, in the southern hemisphere, so the
sense of circulation for an anticyclone is reversed.  So we can take the
$\beta=0.6$ solution in Figure \ref{fig:beta2} and reverse the direction of
the shear flow and the vortex circulation; the result has a slightly
triangular shape pointing towards the equator with relatively rapid westward
flow around the north side.  These features can be seen in the Red Spot.

\subsection{6.B. Comparison with continuous case}

For the continuous version (see \cite{Swaminathan}), with $\beta=0$, DBSA is used
directly on the original equations \eqref{upper}  and \eqref{pv175} using a $512^2$
doubly-periodic, pseudospectral QG model.  The initial condition has sinusoidal
zonal flows and a vortex represented by
$$
q'=\exp(-[r/r_0]^{4})\,.
$$ 
Because of the periodicity, the constraints are tapered to make them
periodic at the boundaries, e.g., 
$$
C_2=\int_{-3\pi}^{3\pi} d\x\; q'\; x\left[1-e^{5(x-3\pi)}\right]
\left[1-e^{5(x+3\pi)}\right]
$$ 
but otherwise the approach follows that in Ref.~\cite{pjmF11}.

\begin{figure}[htb]
\centering
\includegraphics[width=12 cm]{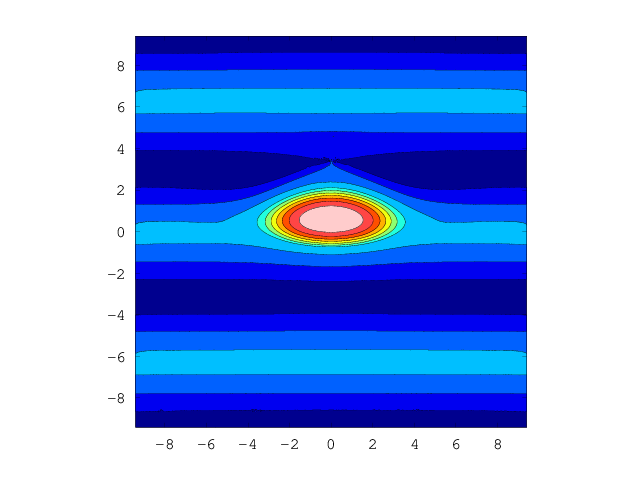}\\
\caption{Potential vorticity $q$ for a vortex centered at $y=1$.  The
translation speed is $c=-0.575$.}
\label{fig:beta0qg}
\end{figure}

The case with the vortex centered is fairly straightforward, and gives
solutions looking very similar to the CD solutions.  Figure \ref{fig:beta0qg} shows  an  
off-centered case with $y_0=1$.  We have used a southern hemisphere situation, consequently, 
the signs of the $\cos(y)$ and $q_0$ terms are reversed.  The
estimated phase speed is $c=-0.575$.

When we include the beta-effect, the standard DBSA procedure works less well,
perhaps because, at least initially, it tries to generate a net north-south
flow which is problematic with the $\beta v$ term.  So we have taken a
somewhat different tack: we solve the modified dynamics problem to find a
steady state ($\alpha=1$)
$$
[\psi_0-\cos y + cy, q_0] = 0
~~~,~~~
q_0 = (\del^2-F+1)\psi_0
$$ with the propagation speed $c=-\int q_0\sin y/\int q_0$.  This can be
rewritten by adding and subtracting $-\cos y+cy$ as
$$
[\psi_0-\cos y + cy, (\del^2-F)\psi_0+\cos y-cy] = 0
$$
or, if we chooses $\beta=-c$ 
\bqy
&&[\psi_0-\cos y , q]= \beta\ppx q
\nonumber\\
&& q =(\del^2-F)\psi_0+\cos y+\beta y] = (\del^2-F)(\psi_0-\cos y) -
F\cos y+\beta y\,.
\nonumber
\eqy
Thus, if we use as an initial condition the $\psi_0$ found from DBSA applied
to the modified dynamics, the resulting structure should propagate at a speed
$c=-\beta$.

\begin{figure}[htb]
\centering
\subfigure[{\footnotesize \  }]{\includegraphics[width=0.48\textwidth]{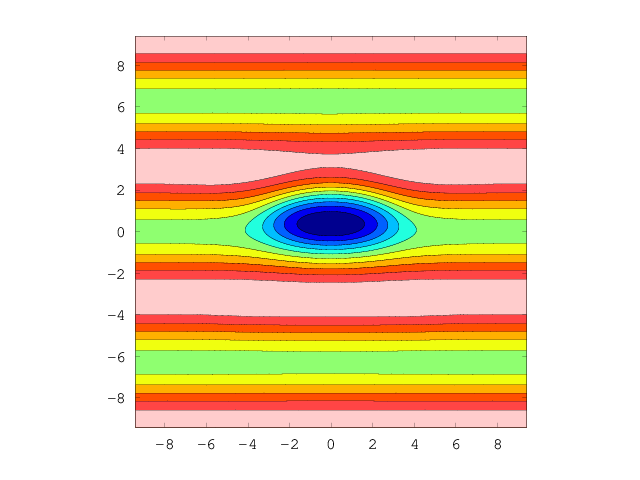}
\label{fig:beta4qg1}
}
\subfigure[{\footnotesize \  }]{\includegraphics[width=0.48\textwidth]{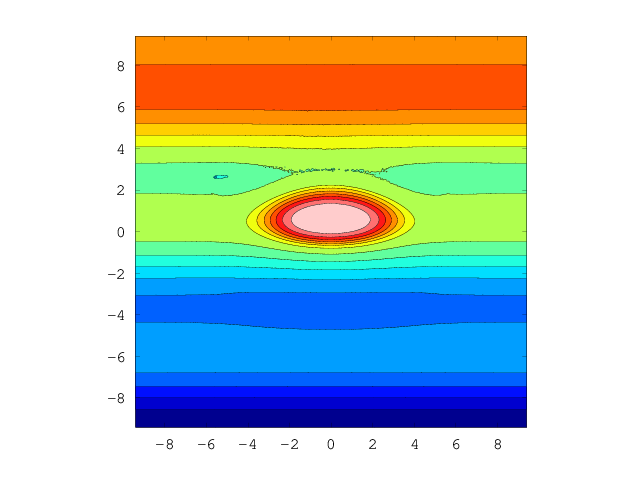}
\label{fig:beta4qg2}
}
\caption[sm] {(a) Streamfunction for a vortex centered at $y=1$.  The translation
speed is $c=-0.5=-\beta$. The vortex is offset in $x$ because it has been
allowed to propagate freely for 120 time units.  (b) Potential vorticity for a vortex centered at $y=1$.  The translation
speed is $c=-0.5=-\beta$. }
\label{fig:beta4qg}
\end{figure}

The $q_0$ field is fairly similar, while the $\psi$ fields, as seen in Figure \ref{fig:beta4qg}  are
virtually indistinguishable.  It shows the rapid flow crossing north of the
vortex, with some more northerly streamlines turning back and merging into the
jet above.  These features, along with the asymmetric bulge to the north, are
noticeable in the movies of the flow near the Red Spot.


\section{7.  Summary and Conclusions}

We have studied the $1\frac{3}{4}$ layer model with deep sinusoidal jets and a
vortex in the upper layer.  The criterion for the stability of the upper layer
jets, which are free to move relative to the deep flow, is that the scale of
the jets (in the sense of the inverse of the wavenumber) must be larger than
the deformation radius.  This is also the necessary condition for isolated
vortices to exist within the upper layer.  In the absence of deep flow, the
jets are not stable and the condition for isolated vortices cannot hold.
Thus, the model argues that the long-lived vortices will be relatively shallow
compared to the jets.

Using a Hamiltonian formulation of contour dynamics and synthetic annealing,
we constructed vortex patch solutions in the sinusoidal flow in the absence of
$\beta$.  These are centered on the line where the shear flow has $u=0$.
Because of the deformation radius and the variations in shear, these are not
precisely elliptical in contrast to the Moore and Saffman \cite{moore} case.

With $\beta$, the shape-preserving vortices will no longer be centered and
will propagate.  To be able to apply the synthetic-annealing procedure, we
introduced a modified form of the dynamics which has the same steady solutions
but evolves according to \eqref{mDyn}.  It should be noted that this procedure
no longer preserves the potential vorticity on each particle as they are
rearranged.  But the final solution is a valid steady state and shows that the
vortex is now off-center and asymmetrical.  It has a triangular component
reminiscent of the Red Spot. 

Physically, there are, of course, other processes acting in the Jovian
atmosphere.  We have used a quasi-geostrophic model, which can become
inaccurate in regions with high Rossby number or, perhaps more appropriate for
large baroclinic vortices, with changes in thickness between isopycnals which
are not small compared to the mean thickness.  The deep layer is not
necessarily completely steady.  This can lead to radiation of waves that drain
energy from the spot; however, estimates of the rate from full two-layer
calculations with the two-beta model \cite{yano94} suggest it is very slow.
Mergers with small spots may spin the large spots back up so that they can be
maintained.  The processes that maintain the deep jets remain unclear, but
convection, moist convection, and baroclinic instability may all play a role.
Despite the many unknowns, we believe that implications of the simple model ---
the shallow spots and the $\beta$-induced asymmetries ---
remain valid.

\bigskip

\noindent{\bf Acknowledgment}.  The authors warmly acknowledge the hospitality of the GFD summer program.  PJM  was supported by  the U.S. Department of Energy Contract DE-FG05-80ET-53088.



\end{document}